# Asymmetric parametric generation of images with nonlinear dielectric metasurfaces

Sergey S. Kruk[1,2#*], Lei Wang[2,3,4#*], Basudeb Sain[1], Zhaogang Dong[5], Joel Yang[5,6],

Thomas Zentgraf[1], and Yuri Kivshar[2*]

[1]Department of Physics, Paderborn University, 33098 Paderborn, Germany
[2]Nonlinear Physics Center, Research School of Physics, Australian National University, Canberra ACT 2601, Australia
[3]National Mobile Communications Research Laboratory, Frontiers Science Center for Mobile Information Communication and Security, Southeast University, Nanjing 210096, China
[4]Purple Mountain Laboratories, Nanjing 211111, China
[5]Institute of Materials Research and Engineering, A*STAR (Agency for Science Technology, and Research), Singapore 138634, Singapore
[6]Singapore University of Technology and Design, Singapore 487372, Singapore

[#]These authors contributed equally to this work

[*]E-mails: sergey.kruk@outlook.com, wang_lei_seu@seu.edu.cn, yuri.kivshar@anu.edu.au

**Subwavelength dielectric resonators assembled into metasurfaces have become a versatile tool for miniaturising optical components approaching the nanoscale[1–3]. An important class of metasurface functionalities is associated with asymmetry in both generation and transmission of light with respect to reversals of the positions of emitters and receivers[4–6]. Nonlinear light-matter interaction in metasurfaces[7–9] offers a promising pathway towards miniaturisation of the asymmetric control of light. Here we demonstrate asymmetric parametric generation of light in nonlinear metasurfaces. We assemble dissimilar nonlinear dielectric resonators into translucent metasurfaces that produce images in the visible spectral range being illuminated by infrared radiation. By design, the metasurfaces produce different and completely independent images for the reversed direction of illumination, that is when the positions of the infrared emitter and the visible light receiver are exchanged. Nonlinearity-enabled asymmetric control of light by subwavelength resonators paves the way towards novel nanophotonic components via dense integration of large quantities of nonlinear resonators into compact metasurface designs.**

The nonlinear optical regime describing interactions of metasurfaces with strong optical fields allow to acquire novel functionalities of metasurfaces beyond those of their linear counterparts[7–9]. Recent examples include imaging through nonlinear meta-lenses [10], nonlinear topological transitions [11], bistability [12], and optical pulse shaping [13]. The efficiency of nonlinear processes in subwavelength photonics can be increased by using high index dielectric nanoparticles supporting multipolar Mie resonances and composite modes[14]. An important cluster of functionalities enabled by nonlinear light-matter interactions is associated with asymmetric control of light [4–6,15,16].

Traditionally, asymmetric control of light is realized with bulky optical components [4–6]. Relatively more compact photonic chips were demonstrated using coupled waveguide micro-resonators [15]. A plasmonic nonlinear metasurface with asymmetries in optical response was demonstrated experimentally [16]. The metasurface functionality relied on a geometric phase gradient created over a distance of several wavelengths. Thus, the approaches experimentally demonstrated to date relied on building blocks that were substantially larger than the wavelength of light in at least one spatial dimension. This hindered the dense integration of large quantities of such photonic components into compact systems analogous to the integration of large quantities of electronic components into semiconductor chips. Recently, there has been an interest in theoretical investigations of various nonlinear photonics platforms for asymmetric light control at the subwavelength scale [17–22]. An earlier theoretical concept of asymmetric nonlinear control of electromagnetic waves with metaphotonics was proposed by A. Mahmoud et. al[17]. The authors presented a nonlinear structure acting as a Faraday rotator without an external magnetic field. For this design, they employed a combination of chiral metasurface elements with strongly nonlinear metasurface elements integrated into a Fabry-Pérot cavity. To achieve a strong nonlinear response, the authors proposed to implement varactor diodes into the design, which made that concept immediately applicable in the microwave frequency range but imposed challenges for its realisation in optics. To this date, none of the existing theoretical proposals for nonlinear asymmetric control of light[17–22] was realized experimentally.

Here we develop theoretically and demonstrate experimentally all-dielectric metasurfaces producing a nonlinear asymmetric generation of light. When infrared light passes through the metasurfaces, certain encoded images are observed, as sketched in Figure 1. However, once we flip the metasurfaces to the opposite side, we observe completely different images.

The functionality of the developed resonators relies on an interplay between nonlinear light-matter interactions and magneto-electric coupling [23,24] to its artificially engineered optical modes. Magneto-electric coupling has facilitated many peculiar photonic functionalities both in microwaves and optics, including polarisation transformations [25,26], anomalous transmission [27,28] and reflection [29,30]; photonic analogues of spin-Hall effects [31], photonic Jackiw-Rebbi states [32], and nontrivial topological phases [33].

We first study metasurfaces made of identical asymmetric nonlinear resonators. The resonators are nano-cylinders whose optical response at the fundamental wavelength is dominated by two Mie multipoles: electric and magnetic dipoles with smaller contributions from higher-order multipoles. The nano-cylinders consist of two layers of materials with different optical constants: amorphous silicon and silicon nitride [see Figs. 2a,b]. Silicon in comparison to silicon nitride has a higher refractive index and higher nonlinear susceptibility. In our geometry, the magneto-electric coupling rises from the asymmetry introduced by the difference in the refractive indices between the two layers. The metasurface is embedded into a homogeneous environment (glass). The geometrical simplicity of our design makes it particularly suitable for standard nanofabrication, such as single-step nanolithography techniques.

Figures 2c,d show results of full-wave simulations of our metasurface unit cell. We calculate the total linear extinction cross-section per-unit-cell $C_{ext}^{total}$ of the nanoresonators of various radii within the metasurface (see details in Methods). In all various radii cases, the resonators feature identical total extinction for "forward" and "backward" illumination as required by reciprocity. However, the anisotropy of the resonators' design reveals itself in a decomposition of the extinction into a series of Mie multipoles $C_{ext}^{total} = C_{ext}^{ED} + C_{ext}^{MD} + C_{ext}^{EQ} + C_{ext}^{MQ} \cdots$ , for simplicity, we refer to the multipoles as ED, MD, et cetera. Figures 2e-h visualize extinction cross-sections of the two dominant multipoles: the electric dipole (ED) and the magnetic dipole (MD) in forward and backward directions. In addition, Supplementary Figure S2 includes information about higher-order multipoles. The Mie resonances interact differently in forward and backward directions which we associate with their magneto-electric coupling. Specifically, the "forward" illumination leads to the enhancement of the magnetic dipole and the suppression of the electric dipole. While the "backward" illumination leads to the opposite effect of the electric dipole enhancement via the magnetic dipole suppression. The modes demonstrate complex dynamics that is rendered by their finite spectral width. We notice that the electric dipole and the magnetic dipole Mie resonances tend to have different spectral width and different Q-factors that is commonly observed in subwavelength dielectric particles[1]. Once higher-order multipoles are considered, the effects of the enhancement/suppression of the resonances are observed between the groups of symmetric and anti-symmetric multipoles dominated by the two dipole modes (details in Supplementary Figure S2). We provide in the Methods section a simplistic analytical insight into the mechanism of enhancement/suppression by the magneto-electric coupling of the two dominant modes, ED and MD. We choose a resonator with 215 nm radius that provides the highest contrast of the MD extinction between the forward and the backward directions. Figures 2i,j show the corresponding multipolar composition. At around 1475 nm wavelength, the resonator featured MD maximum for "forward" illumination and MD minimum for "backward" direction. We note that the effects of the MD enhancement/suppression are inherent properties of our bi-layer particles, and they are also expected from individual, stand-alone nanoresonators (see Supplementary Figures S1a-d and Supplementary Note 1).

Next, we consider the generation of third optical harmonics within the nanoresonators under intense laser illumination. Third harmonic generation (THG) is a process upon which three photons from an excitation beam get "combined" into a single photon with tripled energy and correspondingly tripled frequency. The THG efficiency here depends on optical modes excited within the resonator, and it is drastically higher for the MD mode compared to the ED mode[9]. As such, from the multipolar compositions in Figs. 2i,j we expect brighter THG in "forward" direction (see Fig. 3a). Our analysis to this point ignores optical modes of the resonator at the THG wavelength. We however expect them to have smaller effect on the overall THG efficiency compared to the modes at the pump wavelength[9]. It is worth noting that the metasurface consists of subwavelength unit cells at the fundamental frequency, however the unit cell size becomes larger than the wavelength at the third harmonic frequency.

We proceed with full-wave nonlinear calculations of THG process in our metasurfaces with details presented in Methods. Our numerical simulations consider near-field distributions at both the fundamental and the THG wavelengths thus taking into account optical modes at both frequencies. Figure 3f shows the results of our linear and nonlinear calculations featuring approx. 450 times contrast in THG intensity at around 1475 nm wavelength. At the same time, the metasurfaces have identical linear transmission (Fig. 3c). We find it illustrative to compare THG asymmetric generation in our metasurface with THG in a continuous unpatterned bilayer film with otherwise identical parameters (Figs. 3b,e). The THG contrast of such film does not exceed 1.1. This demonstrates the key role of the interplay of Mie modes in our carefully engineered nanoresonators.

We fabricate the metasurface with electron beam lithography (see details in Methods and an image of a fabricated metasurface in Fig. 3d inset). We measure virtually identical linear transmission of the metasurfaces (Fig. 3d) for "forward" and "backward" scenarios. We next measure the intensity of the third-harmonic signal in the transmission direction for both scenarios of excitation (Fig. 3g) and observe an approx. 25 times difference in the intensity of the third harmonic generation. We note that the contrast in THG is also expected from individual, stand-alone nanoresonators (see Supplementary Figure S1e and Supplementary Note 1).

We proceed from uniform metasurfaces consisting of identical nanoresonators to non-uniform metasurfaces assembled from a set of dissimilar resonators. Such resonators can generate light via nonlinear parametric processes with different parameters, including different intensity, phase, and polarization. We optimise the resonators to produce different levels of THG intensity. With this, we aim to generate completely different images for "forward" and "backward" excitations. We limit ourselves to images with binary intensity. This requires at least four different types of nanoresonators: an *always bright* nanoresonator, an *always dark* nanoresonator, a nanoresonator that is *bright for the "forward", but is dark for the "backward"* directions, and finally a nanoresonator with the opposite functionality that is *dark for "forward" but is bright for "backward"*.

For the sake of nanofabrication simplicity, we require that all the resonators have the same thickness of silicon and silicon nitride layers. Therefore, to add flexibility to the design approach, we introduce two additional geometrical degrees of freedom. We place the resonators on a glass substrate in air. Thus, the resonators are no longer embedded into a homogeneous environment. The presence of a substrate introduces an additional asymmetry to the system which may also affect the third harmonic generation process via the magneto-electric coupling [24]. Second, we allow cylinders with elliptical cross-sections. An extra geometrical parameter (ellipticity of the cross-section) allows us to equalise the relative brightness levels of the four resonators. To avoid additional polarization effects due to the ellipticity of the resonators, we perform all our studies for the same polarization fixed along one of the axes of the ellipses. Supplementary Figure S3 contains details of numerical optimization of the set of four resonators.

Figure 4 (top) shows the design of the four resonators marked α, β, γ, δ and their corresponding theoretically calculated nonlinear optical responses to the "forward" and "backward" directions of excitation (middle). The calculated conversion coefficient $P^{3\omega}/(P^{\omega})^3$ reaches $5 \cdot 10^{-7}$ [W$^{-2}$] for α pillar and backward illumination. If we assume power density of 1GW/cm$^2$ that would give the conversion efficiency $P^{3\omega}/P^{\omega}$ exceeding $3 \cdot 10^{-5}$ (see details in Supplementary Note 3 and Supplementary Tables S1-S3). The theoretical design features at least 9 times THG intensity contrast in the forward direction and at least 8 times contrast in the backward direction.

Figures 5a-b show electron micrographs of three different metasurfaces assembled from the set of four resonators. The metasurface featured in Fig. 5a consists of only two types of nanoresonators (α & β). As shown in Figure 5d (left side), in the "forward" direction, the metasurface is designed to generate a bright square, and in the "backward" direction, it features a bright circle with an embedded dark square. The right side of Fig. 5d shows experimentally observed distributions of third harmonic signals across the metasurface as seen in the transmission direction. The experimental observations confirm our theoretical design. Figures 5b,e feature a metasurface consisting of all four types of nanoresonators and producing a simple independent pair of images: an image of stripes (forward) and a chess-board pattern (backward). We use the images of the stripes and the chessboard pattern to estimate experimental THG conversion efficiencies of the four dissimilar quadrants in both forward and backward directions (see details in Supplementary Note 4 and Supplementary Tables S4-S8). Our estimate suggests that in experiments the conversion coefficienct for α pillar and backward illumination is between $1.5 \cdot 10^{-8}$ & $7 \cdot 10^{-8}$ [W$^{-2}$] which is an order of magnitude lower than the theoretical estimate. We attribute this to fabrication imperfections. The experimental images feature at least 4.5 times contrast in the forward direction and at least 4 times contrast in the backward direction.

Figures 5c,f feature a metasurface with encoded pair of arbitrarily images: a contour of Australia and a stylistic image of a Sydney Opera House. Experimentally observed field distributions resemble closely encoded images. We note that in Figs. 5c,f the edges between the domains of dissimilar resonators can be observed. These artefacts arise because near the edges of the domains, the coupling conditions between the near neighbours alter causing alternation of the THG brightness. In Supplementary Figure S4 and Supplementary Note 2, we consider one example of an artefact in THG brightness occurring near the edge between the two domains. We demonstrate that a slight alternation of sizes of resonators near the domain wall (by 5 nm in radius, or by approx.1.5% in the geometrical cross-section) mitigates the artefact. This example suggests that the image imperfections near the edges can be suppressed via a slight optimization of the design of the near-edge resonators.

In conclusion, we have demonstrated flexible engineering of asymmetric harmonic generation with metasurfaces. Our metasurfaces are translucent slides structured at the nanoscale that generate different images at the third harmonic wavelength when illuminated from the opposite sides. This functionality is enabled by artificially engineered optical interactions beyond the usual electric dipole response, specifically by the optically induced magnetic response and magneto-electric coupling. Our findings pave the way towards novel functionalities of nanoscale optical devices beyond the limits of linear optics. We envision the development of subwavelength resonators and metasurfaces governed by the interplay between magneto-electric coupling and nonlinear light-matter interactions for asymmetric control over multiple parameters of the electromagnetic space, including spatially modulated amplitude, phase, and polarization of optical beams. Functionalities of such resonators can go beyond the parametric generation of light, and the principles revealed in this work may find applications in asymmetric generation of entangled photon states as well as in asymmetric self-action effects leading to nonreciprocity and optical isolation at the nanoscale.

## ACKNOWLEDGEMENTS

The authors thank V. Asadchy, A. Alu, C. Caloz, A. Poddubny, D. Smirnova, K. Simovski, I. Shadrivov, and S. Tretyakov for numerous stimulating discussions. The authors acknowledge the use of the nanofabrication facility at the Paderborn University and acknowledge the Australian National Fabrication Facility, ACT Node, for access to the electron microscope. S.K. acknowledges support from the Alexander von Humboldt Foundation, the Australian Research Council (DE210100679). This project has received funding from the European Union's Horizon 2020 research and innovation programme under the Marie Sklodowska-Curie grant agreement No. : 896735. Z.D. acknowledges help from F. Tjiptoharsono with the fabrication etching recipe. T.Z. acknowledges funding by the European Research Council (ERC) under the European Union's Horizon 2020 research and innovation program (grant agreement No. 724306) and the Deutsche Forschungsgemeinschaft (DFG, German Research Foundation) – TRR142 – No. 231447078 – project B09. L.W acknowledges supports from National Key R&D Program of China (2020YFB1806603), National Natural Science Foundation of China (Grant No. 62101127), Natural Science Foundation of Jiangsu Province of China (BK20200393), SC project of Jiangsu Province (JSSCBS20210116), the Fundamental Research Funds for the Central Universities

(2242022R10025). Y.K. acknowledges a support from the Strategic Fund of the Australian National University, the Australian Research Council (grant DP210101292), and the

## AUTHOR CONTRIBUTIONS

S.K. and L.W. conceived the idea. L.W. performed theoretical calculations. B.S., Z.D., and S.K. fabricated the samples. S.K. and L.W. performed experimental measurements. J.Y., T.Z., and Y.K. contributed to data analysis and to supervision of the project. S.K. wrote the first version of the article. All co-authors contributed extensively to writing and to revisions of the manuscript.

## COMPETING INTEREST STATEMENT

The authors declare no competing interests

## FIGURES

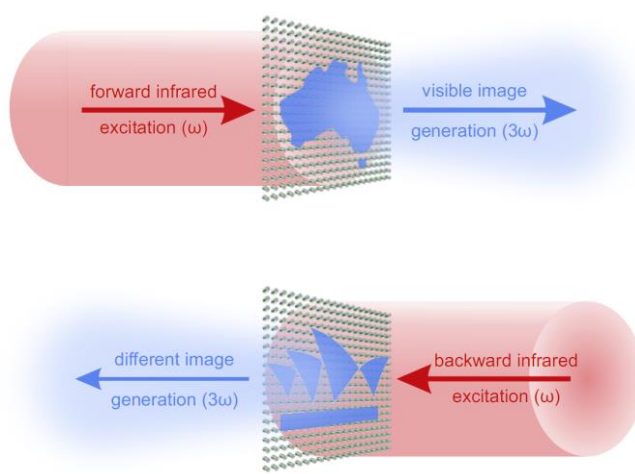

**Figure 1. Concept of asymmetric parametric generation of images with nonlinear metasurface.** The metasurface generates different and independent images in transmission for two opposite illumination directions. The images are produced at tripled frequency via a nonlinear process of third-harmonic generation.

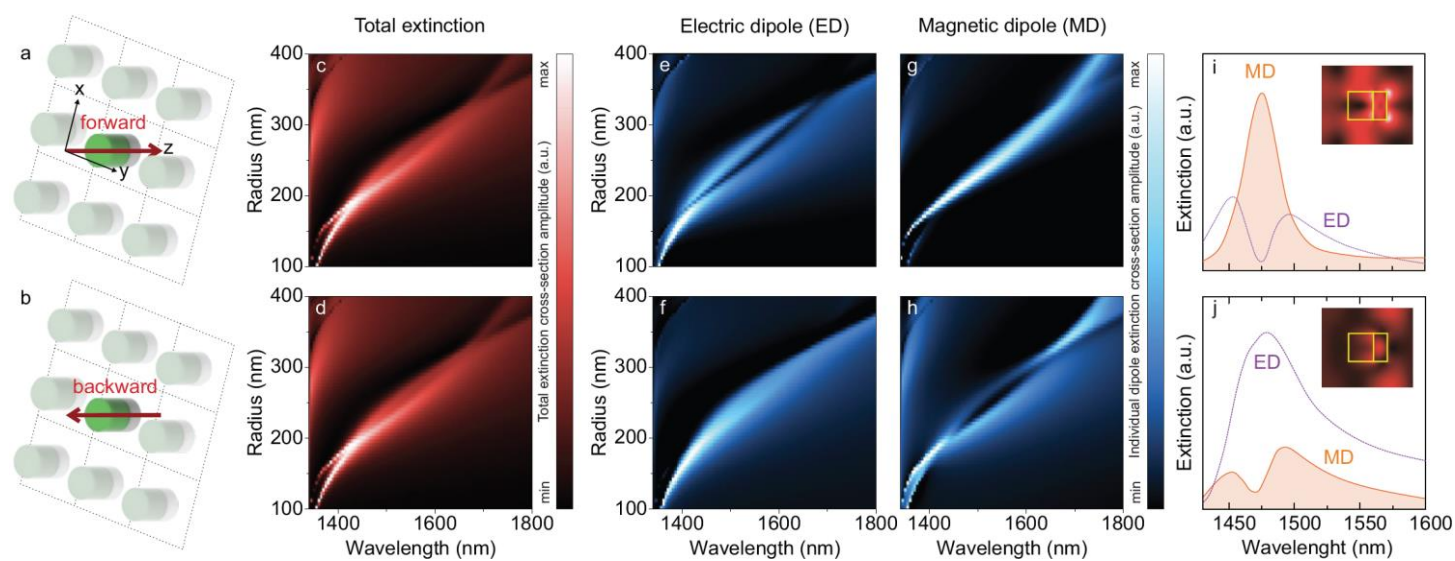

**Figure 2. Optical properties of a metasurface unit cell.** (a,b) An anisotropic cylindrical nanoresonator consisting of two materials: silicon (grey) 220 nm thick and silicon nitride (green) 400 nm thick embedded into glass. The nanoresonator is placed in a square lattice of identical neighbours with a 925 nm period forming a metasurface. Arrows visualise forward (a) and backward (b) directions of excitation. (c,d) Linear extinction spectra of the nanoresonators with 100-400 nm radii in the lattice for forward (c) and backward (d) excitations. (e-h) Contribution of the electric dipole ED (e,f) and the magnetic dipole MD (g,h) for forward (e,g) and backward (f,h) directions. (i,j) Spectra for the electric dipole ED (dashed purple curve) and the magnetic dipole MD (orange filled) of the resonator with 215 nm radius in forward (i) and backward (j) directions. The insets in (i,j) show the nearfield distributions of the electric field amplitude for an excitation wavelength of 1475 nm for (i) forward and (j) backward directions. Yellow lines in the insets mark the contours of the bilayer nanoresonator.

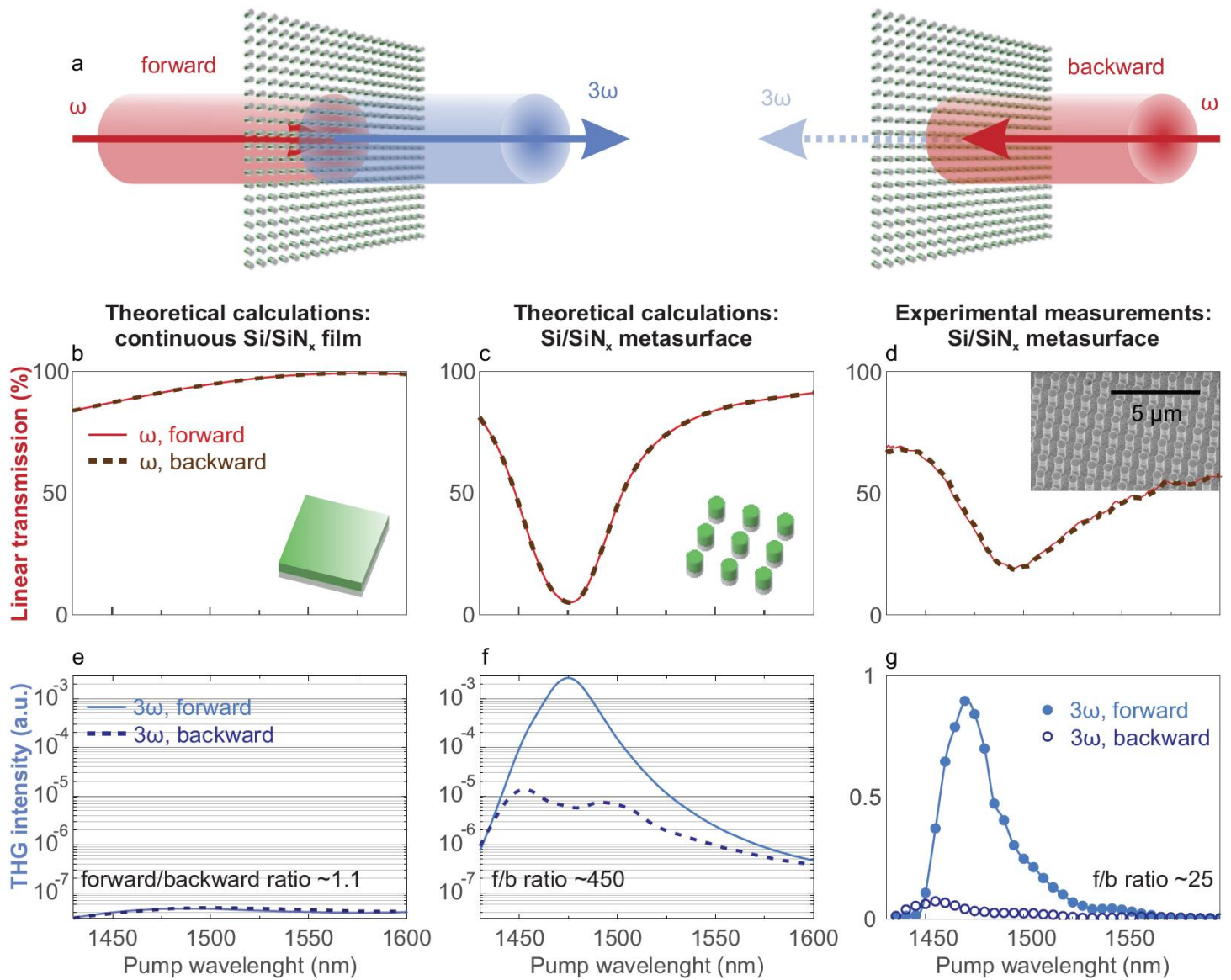

**Figure 3. Symmetric *linear* and asymmetric *nonlinear* spectral response of a uniform metasurface.** (a) Concept image of two scenarios of third harmonic generation: for forward and backward directions of excitation. (b-d) linear transmission and (e-f) third harmonic generation for forward and backward excitation. (b,e) Theoretically calculated response of a continuous unpatterned bi-layer film of silicon (220 nm) and silicon nitride (400 nm) surrounded by glass. (c,f) Theoretically calculated response of a metasurface consisting of silicon and silicon nitride layers of the same thickness (220 nm and 400 nm correspondingly) embedded into glass, cylinder radius 215 nm, square lattice period 925 nm. (d,g) Experimental realization of the bi-layer metasurface. Inset: electron microscope image of the sample (before its embedding into the homogeneous environment)

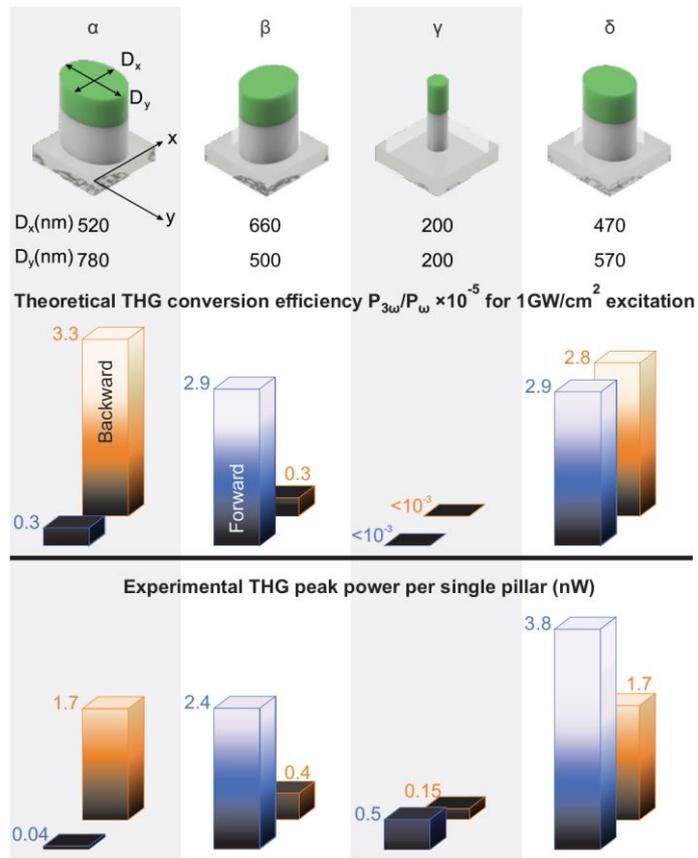

**Figure 4. Set of resonators with dissimilar asymmetric parametric generation of light.** Set of four resonators (α-δ) on a glass substrate in air. All the resonators consist of silicon (grey) layer 480 nm thick and silicon nitride layer (green) 360 nm thick. All the resonators are arranged into a square lattice with a 900 nm period. Elliptical geometries of the resonators in the XY plane are listed in the figure. The α-δ resonators produce four different combinations of THG responses for forward and backward illumination at 1475 nm wavelength as specified in the bar charts.

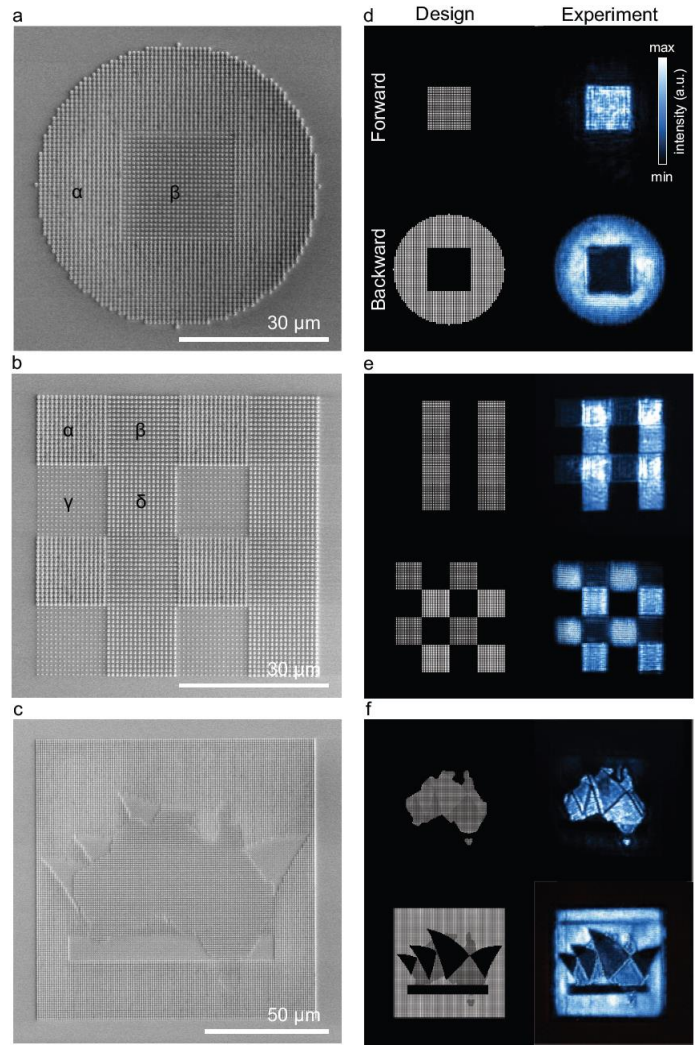

**Figure 5. Asymmetric parametric generation of images with nonlinear metasurfaces.** (a-c) Electron microscope images of three different metasurfaces layouts assembled from the set of four resonators. (d-f) Their nonlinear optical response was detected in transmission at the third harmonic frequency for forward and backward excitation at 1475 nm wavelength. Each of the six experimental images was assigned its individual min. and max. levels of camera counts.

## CODE AVAILABILITY

The code used for modelling the data is available for download at
https://pan.seu.edu.cn:443/link/4A3683E9DA843E10D06FFBA5B43DCDFD

Additional information will be provided by L.W. on reasonable request.

## DATA AVAILABILITY

All data in this study are available within the paper and the supplementary material. Additional information will be provided by S.K. and L.W. on reasonable request.

## METHODS

**Numerical calculations.** We simulate the generation of the third harmonic signal using the finite-element-method solver in COMSOL multiphysics in the frequency domain. All our simulations model the metasurfaces as periodic structures with identical nanoresonators with floquetboundary conditions. Our nonlinear calculations are based on undepleted pump approximation and are performed in two stages. Upon the first stage, we calculate linear light-matter interactions and retrieve electronmagnetic field distributions at the fundamental frequency in the nanoresonator. The pump is a plane wave with power density of 1GW/cm$^2$. We then deduce the current density $J_{THG}$ of the THG within the silicon particle by multiplying the third-order nonlinear susceptibility tensor $\chi^{(3)}$ and the linear field distributions. At the second stage, we employ $J_{THG}$ as a source of external current density for the simulation and retrieve the nonlinear optic field distribution at THG frequency. We approximate the $\chi^{(3)}$ tensor for amorphous silicon to a non-dispersive scalar value of $2.45 \times 10^{-19}$ m$^2$/V$^2$. For the case of metasurface composed of dissimilar nanoresonators, we simplify the model by assuming the coupling between non-identical nanoresonators is roughly the same as that with identical ones.

To find an optimal design for the nanoresonators, we optimize the parameter space of five geometrical dimensions: thickness of each material layer (h1, h2), long and short axes of ellipses (D1, D2), and the square unit cell size (P) (This is reduced to four parameter space with D1=D2 for the homogeneous metasurface).

To enhance the computing speed, we implement the multigoal global genetic optimization algorithms to guide the COMSOL calculations. This method analyses the calculated data sets in each generation as goal functions with the corresponding selecting criteria depending on the desirable characteristics: E.g., selecting criteria for metasurface in Fig2 are (s1) & (s2). For atoms in Fig4 they are α:(s1) & (s3), β: (s1) & (s4), γ: (s6), δ:(s1) & (s5), respectively. The criteria are (s1): Third harmonic conversion efficiency in the desirable illumination direction. (s2): Third harmonic intensity contrast between the forward and backward illumination regimes. (s3): THG predominantly in the direction of the incident wave for backward cases of illumination. (s4): THG for forward cases of illumination. (s5): THG for both cases of illumination. (s6): Low THG efficiency for both illuminations. The geometrical parameters set of the nano-resonators are treated as genomes in this algorithm that evolve to reach optimization. Supplementary Figure S3 illustrates the results of the numerical investigation.

**Analytical description of the magneto-electric coupling.** To yield a simplistic analytical insight, we approximate the optical response of our nanoresonators down to only electric and magnetic dipole modes leaving weaker higher-order multipoles without consideration. We assume that the metasurface response can be grasped by effective polarizability tensors $\bar{\bar{\alpha}}_{ee}$ (electric), $\bar{\bar{\alpha}}_{mm}$ (magnetic), and $\bar{\bar{\alpha}}_{me}$ (magnetoelectric). Without a lack of generality, we will focus on x-polarized light beams propagating in $\pm z$ directions. The response will thus be governed by $\alpha_{ee}^{xx}$, $\alpha_{mm}^{yy}$, $\alpha_{me}^{xy}$, and $\alpha_{me}^{yx}$ components of the tensors. We further assume that the metasurface obeys optical reciprocity in the linear regime. This further reduces the number of polarizability terms: $\alpha_{me}^{xy} = -\alpha_{me}^{yx}$. Under this condition, the excitation of the electric and the magnetic dipole modes can be written as:

$$p_x = \alpha_{ee}^{xx} E_x \pm \alpha_{me}^{xy} H_y$$
$$m_y = \pm \alpha_{mm}^{yy} H_y - \alpha_{me}^{xy} E_x$$

(1)

where $E_x$ and $H_y$ are local fields at the position of a nanoresonator. The "$\pm$" sign in front of the "$H_y$" term accounts for the cases of forward/backward illumination. From Eqs. 1 we can see that the magnetic dipole mode is enhanced for the forward illumination, and it is suppressed for the backward illumination by the magneto-electric coupling. The enhancement/suppression of the electric dipole mode is the opposite of that of the magnetic dipole. Notably, the strength of magneto-electric coupling described by $\alpha_{me}^{xy}$ conceptually may take arbitrary large values, reaching or even exceeding the strengths of electric and magnetic responses described via $\alpha_{ee}^{xx}$ and $\alpha_{mm}^{yy}$ components[34,35]. Interestingly, a balance between the polarizability tensors may lead to complete suppression of a given dipole mode for a given direction of excitation:

$$p_x = 0 \quad \text{for} \quad Z_0 \alpha_{ee}^{xx} = \alpha_{me}^{xy} \quad \text{and forward} \ (+H_y) \ \text{propagation}$$
$$m_y = 0 \quad \text{for} \quad \frac{1}{Z_0} \alpha_{mm}^{yy} = \alpha_{me}^{xy} \quad \text{and backward} \ (-H_y) \ \text{propagation}$$
(2)

Here $Z_0$ is the impedance of the homogeneous environment.

**Calculation of extinction and multipole coefficients.** We derive the total extinction cross-section per-unit-cell by evaluating the energy flux and total loss within the unit cell through numerically solved near-field information. Total extinction is the sum of total scattering and absorption $C_{ext}^{total} = C_{scatter}^{total} + C_{absorb}^{total}$. The total scattering cross-section per-unit-cell $C_{scatter}^{total}$ is equal to $Q_{sca}/I_0$, the scattered energy flux $Q_{sca}$ can be computed via surface integral of Poynting vector of scattered field over the open boundaries of the unit cell, and $I_0$ is the intensity of the incident light wave. $C_{absorb}^{total}$ is equal to the volume integral of the loss $Q_h$ in the domain of the nanoresonator, which is usually neglectable for silicon material.

The electric and magnetic multipole coefficients can be derived from the numerically solved scattering current density $\mathbf{J}(\mathbf{r})$ by solving partial differential equations using the multipole decomposition methods detailed in [36]. The multipole terms for averaged extinction per-unit-cell $C_{ext}^{ED}, C_{ext}^{MD}, C_{ext}^{EQ}, C_{ext}^{MQ} \cdots$, can therefore be calculated using Eqs.3 with $l = 1,2,3$ as dipole, quadrupole and octupole terms.

$$C_{ext}^{E,l} = -\frac{\pi}{k^2} \Sigma_{m=-1,+1}(2l+1) Re[ma_E(l,m)]$$
$$C_{ext}^{M,l} = -\frac{\pi}{k^2} \Sigma_{m=-1,+1}(2l+1) Re[a_M(l,m)]$$
(3)

In this work, the multipolar analysis was performed using COMSOL. Some of the technical details on multipole analysis in COMSOL can be found in Ref. [37]

**Nanofabrication.** The metasurfaces were fabricated on a glass substrate using a multi-step process that begins with consecutive deposition of thin films of amorphous silicon (a-Si) and silicon nitride (SiN$_x$) of desired thicknesses by plasma-enhanced chemical vapor deposition (PECVD). For the homogeneous metasurface (Fig. 3), a silicon nitride thin film was deposited first followed by the deposition of an amorphous silicon film, while for the inhomogeneous metasurface (Fig. 4), the silicon nitride film was deposited after the amorphous silicon film. A poly-methyl-methacrylate (PMMA) resist layer was spin-coated onto the bilayer films and baked on a hot plate at 170 °C for 2 min to remove the solvent. Then, the desired patterns were transferred by using a standard electron beam lithography and subsequent development in 1:3 methyl isobutyl ketone (MIBK): isopropyl alcohol (IPA) solution. During the electron beam writing process, multiple sets of samples were patterned with both positive and negative size biases to accommodate for possible size deviations during the fabrication. Next, a 20 nm thick Chromium (Cr) mask was deposited by electron beam evaporation. After a lift-off process in hot acetone, the patterns were transferred from PMMA to Cr. Finally, the structures were transferred onto the bilayer material using an inductively coupled plasma reactive ion etching (ICP-RIE) and the subsequent removal of the Cr mask by a commercially purchased Cr-etch solution. Note, each material layer within the bilayer film was etched by a recipe optimised to that specific material. Electron microscope images of the resulting sample were obtained using a scanning electron microscope. The homogeneous metasurfaces were further embedded into a droplet of optical oil with a refractive index like that of the glass substrate and covered by a glass slide identical to the substrate slide.

**Optical measurements.** For linear spectral measurements, a tungsten halogen light bulb was used as a light source. For nonlinear optical measurements, a pulsed laser system was used in the experiments as a light source. It consisted of Femtolux Ekspla laser (1030 nm wavelength) and Hotlight Systems MIROPA optical parametric amplifier (1350-1750 nm wavelengths tunability range). Optical pulses with 2 ps duration, 5MHz repetition rate, and linear horizontal polarization were used. The average output power of the laser system was 500 mW at 1450 nm wavelength decreasing for other wavelengths. The power was monitored by an Ophir IR power meter. The collimated laser beam was narrowed down by achromatic doublet lenses from Thorlabs to illuminate an area of approximately 200±25 µm, which is about twice as large as the metasurface in Fig. 5c, and approximately four times larger than the metasurfaces in Figs. 5a,b. Both the infrared (IR) excitation and the visible (VIS) generated third-harmonic signal was captured by an objective lens Mitutoyo Plan Apo NIR HR X100 NA0.7 with an achromatic performance across 400-1800 nm spectral range. The IR spectra were measured with a spectrometer NIR-Quest by Ocean Optics, and VIS spectra were monitored by a spectrometer QE Pro by Ocean Optics. The metasurface images in the IR were recorded on a camera Xenics Bobcat 320 paired with an infinity-corrected f=150 mm IR achromatic doublet lens from Thorlabs. The metasurface images in the VIS were recorded on a camera Trius-SX694 Starlight Xpress paired with an infinity-corrected f=150 mm VIS achromatic doublet lens from Thorlabs. For measurements at the third-harmonic frequency, the excitation wavelength was filtered by a colour glass FGB900. "Forward" and "Backward" illumination directions were recorded by flipping the sample around the vertical axis. Parameters of the excitation laser were kept the same for "Forward" and "Backward" experiments. Experimental "Backward" images were correspondingly mirrored along the vertical axis in post-processing. An aperture diaphragm was added in the back focal plane of the imaging objective ensuring the collection of only the forward-propagating signal. The resolution of "forward" images is reduced in comparison to "backward" images by aberrations introduced by the substrate.

**METHODS REFERENCES**